\documentclass[showpacs,printnumbers,amsmath,amssymb,twocolumn, aps, pra, longbibliography, superscriptaddress]{revtex4-1}
\usepackage{graphicx}
\usepackage{dcolumn}	
\usepackage{bm}			
\usepackage{amsfonts}
\usepackage{xspace}
\usepackage{color}
\usepackage{epstopdf}

\begin{document}


\newcommand{\psihat}{\ensuremath{\hat{\psi}}\xspace}
\newcommand{\psihatd}{\ensuremath{\hat{\psi}^{\dagger}}\xspace}
\newcommand{\ahat}{\ensuremath{\hat{a}}\xspace}
\newcommand{\Ham}{\ensuremath{\mathcal{H}}\xspace}
\newcommand{\ahatd}{\ensuremath{\hat{a}^{\dagger}}\xspace}
\newcommand{\bhat}{\ensuremath{\hat{b}}\xspace}
\newcommand{\bhatd}{\ensuremath{\hat{b}^{\dagger}}\xspace}
\newcommand{\boldr}{\ensuremath{\mathbf{r}}\xspace}
\newcommand{\dr}{\ensuremath{\,d^3\mathbf{r}}\xspace}
\newcommand{\dk}{\ensuremath{\,d^3\mathbf{k}}\xspace}
\newcommand{\etal}{\emph{et al.\/}\xspace}
\newcommand{\ie}{i.e.}
\newcommand{\eq}[1]{Eq.~(\ref{#1})\xspace}
\newcommand{\fig}[1]{Fig.~\ref{#1}\xspace}
\newcommand{\abs}[1]{\left| #1 \right|}
\newcommand{\proj}[2]{\left| #1 \rangle\langle #2\right| \xspace}
\newcommand{\Qhat}{\ensuremath{\hat{Q}}\xspace}
\newcommand{\Qhatd}{\ensuremath{\hat{Q}^\dag}\xspace}
\newcommand{\phihatd}{\ensuremath{\hat{\phi}^{\dagger}}\xspace}
\newcommand{\phihat}{\ensuremath{\hat{\phi}}\xspace}
\newcommand{\boldk}{\ensuremath{\mathbf{k}}\xspace}
\newcommand{\boldp}{\ensuremath{\mathbf{p}}\xspace}
\newcommand{\boldsigma}{\ensuremath{\boldsymbol\sigma}\xspace}
\newcommand{\boldalpha}{\ensuremath{\boldsymbol\alpha}\xspace}
\newcommand{\grad}{\ensuremath{\boldsymbol\nabla}\xspace}
\newcommand{\parti}[2]{\frac{ \partial #1}{\partial #2} \xspace}
 \newcommand{\vs}[1]{\ensuremath{\boldsymbol{#1}}\xspace}
\renewcommand{\v}[1]{\ensuremath{\mathbf{#1}}\xspace}
\newcommand{\Psihat}{\ensuremath{\hat{\Psi}}\xspace}
\newcommand{\Psihatd}{\ensuremath{\hat{\Psi}^{\dagger}}\xspace}
\newcommand{\Vhatd}{\ensuremath{\hat{V}^{\dagger}}\xspace}
\newcommand{\Xhat}{\ensuremath{\hat{X}}\xspace}
\newcommand{\Xhatd}{\ensuremath{\hat{X}^{\dag}}\xspace}
\newcommand{\Yhat}{\ensuremath{\hat{Y}}\xspace}
\newcommand{\Jhat}{\ensuremath{\hat{J}}\xspace}
\newcommand{\Yhatd}{\ensuremath{\hat{Y}^{\dag}}\xspace}
\newcommand{\jhat}{\ensuremath{\hat{J}}\xspace}
\newcommand{\lhat}{\ensuremath{\hat{L}}\xspace}
\newcommand{\Nhat}{\ensuremath{\hat{N}}\xspace}
\newcommand{\rhohat}{\ensuremath{\hat{\rho}}\xspace}
\newcommand{\ddt}{\ensuremath{\frac{d}{dt}}\xspace}
\newcommand{\nset}{\ensuremath{n_1, n_2,\dots, n_k}\xspace}
\newcommand{\Var}{\ensuremath{\mbox{Var}}\xspace}
\newcommand{\notes}[1]{{\color{blue}#1}}
\newcommand{\cmc}[1]{{\color{red}#1}}
\newcommand{\sah}[1]{{\color{magenta}#1}}

\title{Quantum Noise in Bright Soliton Matterwave Interferometry}
\author{Simon A.~Haine}
\affiliation{Department of Physics and Astronomy, University of Sussex, Brighton BN1 9QH, United Kingdom}
\email{simon.a.haine@gmail.com}

\begin{abstract}
There has been considerable recent interest in matterwave interferometry with bright solitons in quantum gases with attractive interactions, for applications such as rotation sensing. We model the quantum dynamics of these systems and find that the attractive interactions required for the presence of bright solitons causes quantum phase-diffusion, which severely impairs the sensitivity. We propose a scheme that partially restores the sensitivity, but find that in the case of rotation sensing, it is still better to work in a regime with minimal interactions if possible.  
\end{abstract}

\maketitle

\section{Introduction}  Rotation sensors based on matterwave interferometers have the potential to provide state of the art sensing capabilities \cite{Cronin:2009, Riehle:1991}. The current pursuits towards fulfilling this potential can be divided into two main approaches: \emph{Free-space} atom interferometers, which operate in free-fall and use optical transitions between momentum modes to achieve spatial path separation \cite{Gustavson:1997, Lenef:1997, Gustavson:2000, Durfee:2006, Canuel:2006, Dickerson:2013}, or \emph{guided} configurations which involves the propagation of atoms along some guiding potential to achieve spatial path separation, analogous to an optical fibre \cite{GarridoAlzar:2012, Halkyard:2010, Kandes:2013, Stevenson:2015, Kolar:2015, Nolan:2016, Bell:2016, Haine:2016b, Navez:2016}. While both approaches have their advantages, one attraction towards guided configurations is the potential for a very large enclosed area \cite{Stevenson:2015}, and therefore higher per-particle sensitivity. However, guided matterwave interferometery often requires working in a regime where atom-atom interactions are important, leading to complications in the matterwave dynamics \cite{Kandes:2013, Haine:2016b, McDonald:2014, Kolar:2015}. One approach to minimize these effects is to work with atomic gases with attractive interaction in the soliton regime \cite{Strecker:2002, Khaykovich:2002, Strecker:2003, Negretti:2004, Cornish:2006, Veretenov:2007, Kartashov:2011, Martin:2012, Helm:2012, Marchant:2013, Polo:2013, Cuevas:2013, Helm:2014, Helm:2015, Hidetsugu:2016}. In fact, it has recently been shown that soliton interferometry can provide higher fringe contrast than non-interacting gases \cite{Negretti:2004, Veretenov:2007, Helm:2012, Martin:2012, Helm:2014, Helm:2015, Hidetsugu:2016}, although studies that include quantum noise have cast doubt on this increased fringe contrast \cite{Martin:2012, Helm:2014}. In this work, we use the Quantum Fisher Information (QFI) to confirm this suspicion and show that this increased fringe contrast is an artefact of the mean-field model, and that to quantitatively evaluate the sensitivity of bright soliton matterwave interferometry schemes, it is crucial to include the effects of quantum noise. We consider the example of a matterwave gyroscope in a ring-trap, and show that in the case of a non-interacting gas the sensitivity is independent of the shape of the wave-packet. When adding an attractive nonlinearity required for bright solitons, we find that the quantum noise severely degrades the sensitivity. Finally, we find that for intermediate interaction strengths, a modification to the scheme to include the addition of a state-preparation step can partially recover the sensitivity, but argue that it is usually better to minimise the interactions if possible, rather than working in the soliton regime. 

\begin{figure*}[t]
\centering 
\includegraphics[width =0.8\textwidth]{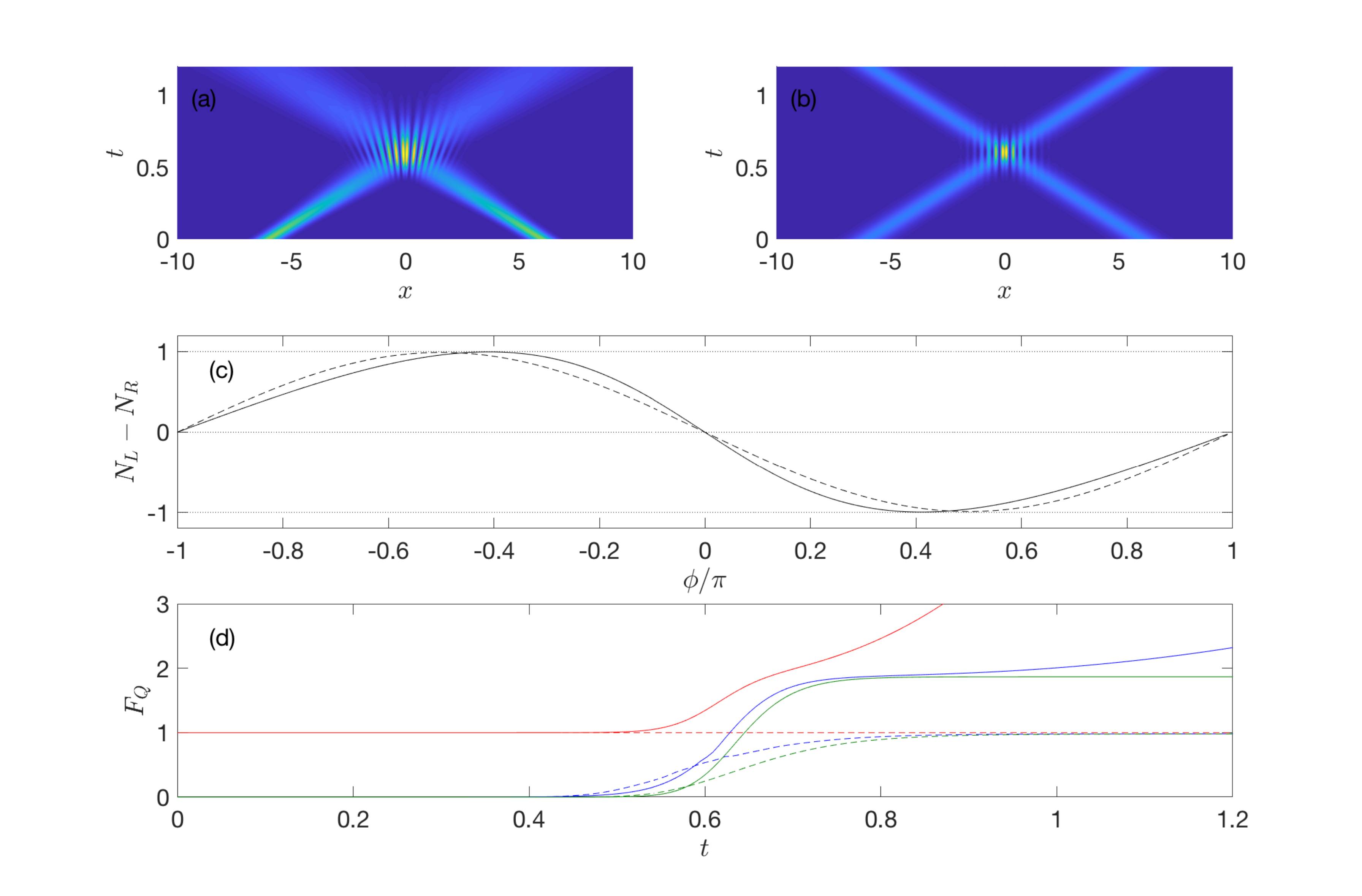}
\caption{GPE simulation of two wave-packets interfering after scattering off a narrow barrier repulsive barrier $V(x) = V_0 \exp (-x^2/w^2)/(w\sqrt{\pi})$. In each case, the initial state was chose as $\Psi(x,0) = \frac{1}{\sqrt{2}}(\Psi_L e^{ikx} + e^{i\phi} \Psi_R e^{-ikx})$. For case 1, we have used a noninteracting system ($g_0 =0$), and a Gaussian wavepacket: $\Psi_{L(R)} = B \exp(-(x-(+) x_0)^2/\sigma^2)$, and for case 2 we have used attractive interactions $g_0 <0$ and a solitonic  initial wavepacket: $\Psi_{L(R)} = B \cosh (\sqrt{2 \mu/\hbar^2}(x-(+) x_0))$, where in both cases $B$ is the appropriate normalisation factor.  (a): Probability density of case 1; (b): probability density of case 2; (c): Dashed (solid) line $P_L-P_R$ vs. $\phi$ for case 1(2); (d): Red dashed (solid) line: $F_Q$ for case 1(2); Green dashed (solid) line: $F_C$ for case 1(2); Blue dashed (solid) line: $F_C^x = \int (\partial_\phi |\Psi|^2)^2/|\Psi|^2 dx$, the CFI for full density resolving measurements for case 1(2). Parameters: $\sigma = 0.5$, $k=10$, $w=10^{-2}$, $V_0 = 5.65$. The barrier height $V_0$ was chosen to give $50\%$ reflection for this value of $k$. We are working in units where $\hbar=m=1$.}
\label{fisher}
\end{figure*}

\section{Fisher Information Bounds for Fringe Contrast} 
The Quantum Fisher Information (QFI) describes how much information about a particular parameter is contained in a quantum state, and through the Quantum Cramer-Rao bound (QCRB), provides strict bounds on how precisely that parameter can be estimated through measurements performed on that state \cite{Demkowicz-Dobrzanski:2014}. More precisely, the QCRB states that by making $\mathcal{M}$ measurements on identically prepared systems, the error in estimates of a particular parameter $\Omega$ is bounded by $\Delta \Omega \geq 1/\sqrt{\mathcal{M}\mathcal{F}_Q}$. Consider the situation described in \cite{Martin:2012, Helm:2012, Helm:2014, Hidetsugu:2016}, where a relative phase shift $\phi$ is applied to matterwave wave-packets of equal population before they collide on a narrow barrier, resulting in 50\% transmission and 50\% reflection. If we consider the full $N$-particle quantum state $|\Psi\rangle$, then the state at some later time $t$ can in general be described by $|\Psi(t)\rangle = \hat{U}|\Psi_\phi\rangle$, where $|\Psi_\phi\rangle$ is the state immediately after the application of the phase shift, $\hat{U} =  \exp(-i\hat{\mathcal{H}} t/\hbar)$ and $\hat{\mathcal{H}}$ is the full $N$-particle Hamiltonian which describes the kinetic energy, potential energy, and arbitrary inter-particle interactions. The QFI of the final state is 
\begin{align}
\mathcal{F}_Q[|\Psi(t)\rangle] &= 4\left[\langle \partial_\phi \Psi(t)|\partial_\phi \Psi(t)\rangle - |\langle \Psi(t)|\partial_\phi \Psi(t)\rangle|^2\right] \notag \\
& = 4\left[\langle \partial_\phi \Psi_\phi| \hat{U}^\dag \hat{U} |\partial_\phi \Psi_\phi \rangle - |\langle \Psi_\phi|\hat{U}^\dag \hat{U} |\partial_\phi \Psi_\phi \rangle|^2\right] \notag \\
&= 4\left[\langle \partial_\phi \Psi_\phi |\partial_\phi \Psi_\phi \rangle - |\langle \Psi_\phi |\partial_\phi \Psi_\phi \rangle|^2\right] \nonumber  \\
&= \mathcal{F}_Q[|\Psi_\phi\rangle]
\end{align}
where we have used the fact that $\hat{U}$ is independent of $\phi$, and $\hat{U}^\dag\hat{U}=1$. That is, $\mathcal{F}_Q$ is unchanged by the subsequent evolution. If the many particle quantum state is initially separable, ie, $|\Psi_\phi\rangle = (\ahatd_\psi)^N/\sqrt{N!}|0\rangle$ where $\ahat_\psi = \int \Psi^*(x)\psihat(x) dx$,  where $\psihat(x)$ is the usual bosonic annihilation operator and $\Psi(x)$ is the single-particle wave-function, then it can be shown that $\mathcal{F}_Q = N F_Q$ \cite{Haine:2016b, Kritsotakis:2017}, where 
\begin{equation}
F_Q = 4\left[\int \partial_\phi \Psi^* \partial_\phi \Psi dx - \abs{\int \Psi^* \partial_\phi \Psi dx}^2\right]
\end{equation}
is the \emph{single-particle} QFI. If $\Psi(x,0) = \frac{1}{\sqrt{2}}(\Psi_L + e^{i \phi} \Psi_R)$, where $\Psi_L$ and $\Psi_R$ are orthonormal wave-packets representing the two wave-packets in the initial configuration, its straight forward to show that $F_Q = 1$.

Meanwhile, after the wave-packets interact with the barrier, the Classical Fisher Information (CFI) is related to the probability of detecting a particle on the left (right) side of the barrier $P_{L(R)}$ by $\mathcal{F}_C = N F_C$, where $F_C = \left[(\partial_\phi P_L)^2/P_L + (\partial_\phi P_R)^2/P_R\right]$ is the \emph{single-particle} CFI. However, when the situation is modelled with the Gross-Pitaeveskii equation (GPE), where $\Psi$ evolves according to $i \hbar \partial_t \Psi = (\frac{-\hbar^2}{2m}\partial_x^2  + V(x) )\Psi + g_0 N |\Psi|^2\Psi$, for attractive interactions $g_0 <0$ and $\Psi_{L(R)}$ set as bright soliton solutions, it was found in \cite{Martin:2012, Helm:2012, Helm:2014, Hidetsugu:2016} that at the optimum point (when $P_L = P_R = \frac{1}{2}$),  $(\partial_\phi P_{L(R)})^2$ can be significantly greater than $1$, indicating $F_C > 1$ and therefore a violation of the QCRB: $\mathcal{F}_C \leq \mathcal{F}_Q$ \cite{Demkowicz-Dobrzanski:2014}. Furthermore, as we show in figure (\ref{fisher}), these simulations show that the QFI increases with time, which is unphysical, and therefore these simulations cannot provide meaningful assessments of metrological usefulness. One possibility for this discrepancy is that the GPE can lead to dynamics with a positive Lyapunov exponent, and therefore caution must be applied when determining it's applicability in some cases \cite{Brezinova:2011}. Of course, the QFI can exceed $N$ when there are non-trivial quantum correlations present. However, the creation of such correlations cannot be modelled by the GPE, which is why models that include the quantum noise should be considered when assessing the metrological usefulness of such devices.  

\begin{figure}
\includegraphics[width=1.0\columnwidth]{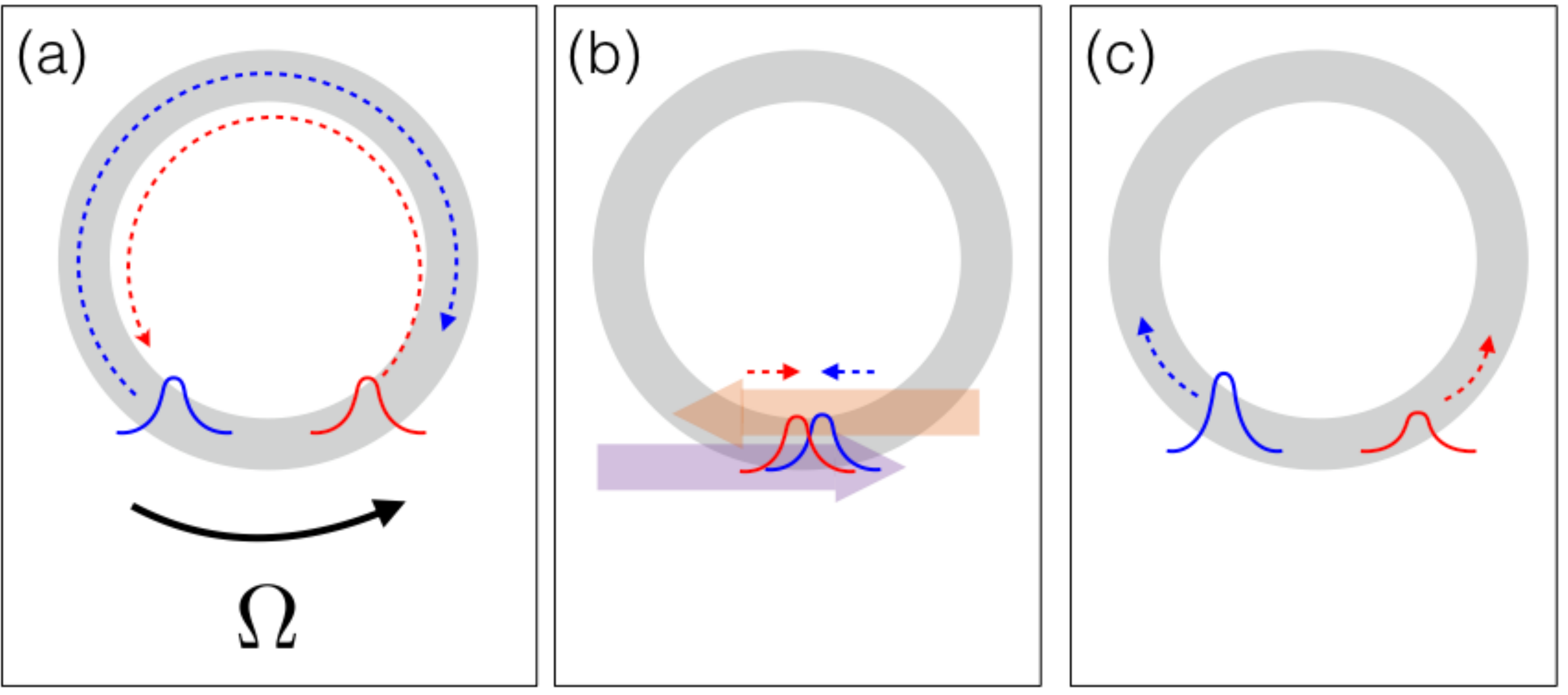}
\caption{(Color Online) (a): Matterwaves formed from a two-component BEC (clockwise: $|+\rangle$, counter-clockwise: $|-\rangle$) propagate around the ring in opposite directions, accumulating a phase-difference due to the external rotation frequency $\Omega$. (b): After one complete circuit, the two components are interfered via a two-photon Raman transition, and the phase difference is converted into population difference (c). }
\label{scheme}
\end{figure}

\section{Matterwave gyroscope} To demonstrate the role of quantum noise, we consider the example of a gyroscope based on interference of matterwaves confined in a ring shaped potential, described in Fig.~(\ref{scheme}). Two counter-propagating matterwaves traverse the ring in opposite directions and are interfered, with the goal of estimating the magnitude of a rotational frequency $\Omega$. We consider a Bose gas consisting of two hyperfine components (electronic states $|+\rangle$ and $|-\rangle$), with bosonic annihilation operators $\psihat_+(\boldr)$ and $\psihat_-(\boldr)$ respectively, which obey they usual bosonic commutation relations: $\left[\psihat_i(\boldr), \psihatd_j(\boldr^\prime)\right] = \delta(\boldr-\boldr^\prime)\delta_{ij}$. An initial state is created with all the atoms in state $|+\rangle$, before implementing an atomic \emph{beamsplitter}, which performs the operation $\psihat_{\pm} \rightarrow \frac{1}{\sqrt{2}}(\psihat_{\pm} \mp \psihat_{\mp} e^{\pm 2i n\vartheta})$, where $\vartheta$ is the angular coordinate around the ring, coherently transferring 50\% of the population to state $|-\rangle$ while also shifting the angular momentum by $-2n\hbar$. Such a process could be implemented via a two-photon Raman transition with Laguerre-Gauss beams as described in \cite{Andersen:2006, Moulder:2012, Halkyard:2010, Nolan:2016}. After time $T = 2\pi R^2 m/\hbar n$, the two components have each traversed the ring and we apply another Raman coupling pulse to act as a second beamsplitter performing the transformation $\psihat_{\pm} \rightarrow \frac{1}{\sqrt{2}}(\psihat_{\pm} -i \psihat_{\mp} e^{\pm 2i n\vartheta})$, before the population in each component is measured and used to infer the phase difference accrued, and therefore estimate $\Omega$. As in \cite{Halkyard:2010, Kandes:2013, Helm:2015, Haine:2016b}, working in cylindrical coordinates, $\{r, \vartheta, z\}$, we assume a trapping potential of the form $V(\boldr) = \frac{1}{2}m\left[ \omega_z^2 z^2 + \omega_r^2 (r-R)^2\right]$ where $R$ is the radius of the torus, $\omega_z$ and $\omega_r$ are the axial and radial trapping frequencies, and $m$ is the mass of the particles. Assuming that the radial and axial confinement is sufficiently tight, we may ignore the dynamics in these directions. In terms of the coordinate $\xi = \vartheta R$,  the effective Hamiltonian for the system is
\begin{eqnarray}
\Ham &=&  \sum_{j = +,-} \int \psihatd_j(\xi) \hat{H}_0 \psihat_j(\xi) d\xi \nonumber \\
 &+& \sum_{i,j = +,-} \frac{g_{ij}}{2} \int \psihatd_j(\xi)\psihatd_i(\xi)\psihat_j(\xi)\psihat_i(\xi) d\xi \, , \label{ham1}
\end{eqnarray}
where $\hat{H}_0 = \frac{-\hbar^2}{2m}\frac{\partial^2}{\partial \xi^2} - \Omega \hat{L}_z$,  and $\hat{L}_z$ is the $z$ component of the angular momentum, and we have assumed that we are working in a frame rotating around the $z$-axis at angular frequency $\Omega$. $g_{ij}$ is the two-particle contact potential interaction strength between state $|i\rangle$ and $|j\rangle$ atoms. For convenience, we assume that $g_{++}= g_{--} \equiv g_0 \leq 0$, and $g_{+-}=0$. The choice of $g_{+-}$ has very little effect on the results as for most of the duration the two components are not spatially overlapping \cite{note1}. 

\subsection{Noninteracting Case} We begin by examining the simple case where $g_0 = 0$, as we can obtain an analytic result with which to benchmark the behaviour in the soliton regime. Working in the Heisenberg picture, and expanding our field operators in angular momentum basis $\psihat_\pm(\xi) = \frac{1}{\sqrt{2\pi}}\sum_{q} \bhat^\pm_q e^{i q \xi/R}$, the operators at some time $t_f$ after the interferometer sequence (beamsplitter/free evolution/beamsplitter) are 
\begin{align}
\bhat_q^\pm(t_f) &= \frac{1}{2}\left[e^{- i \phi_q}(\bhat_q^\pm(0) \mp \bhat_{q\mp 2n}^\mp(0)) \right. \nonumber \\ 
 &- \left. i e^{- i\phi_{q\mp 2n}}(\bhat_q^\pm(0) \pm \bhat_{q \mp 2n}^\mp(0))\right],
\end{align}
where $\phi_q = (\frac{\hbar q^2}{2m R^2} -\Omega q)t_f$. If we use the number difference in each component $\hat{N}_d \equiv \hat{N}_+ - \hat{N}_-$ as our signal, where $\hat{N}_\pm = \int_{-\pi R}^{\pi R} \psihatd_\pm\psihat_\pm d\xi$ then the rotation sensitivity is given by
\begin{equation}
\Delta \Omega = \sqrt{\frac{\Var(\hat{N}_d)}{(\partial_\Omega \langle \hat{N}_d\rangle)^2}} \, . \label{delta_Omega}
\end{equation}
At $t_f = T$, $\phi_{q+2n} - \phi_{q} = \frac{4 m\pi R^2 \Omega}{\hbar} \equiv \phi_\Omega$, where we have subtracted the constant $4\pi(q+n)$ as integer multiples of $2\pi$ are inconsequential. Importantly, $\phi_\Omega$ is independent of $q$, which allows us to greatly simplify $\hat{N}_d$. Assuming $\langle \hat{N}_-(0)\rangle = 0$, we obtain $\Var(\hat{N}_d) = \sin^2 \phi_\Omega \Var(\hat{N}_+(0)) + \cos^2 \phi_\Omega \langle \hat{N}_+(0)\rangle$, and $\langle \hat{N}_d\rangle = \sin\phi_\Omega \langle \hat{N}_+(0)\rangle$. At the most sensitive point $\phi_\Omega = 0$, this simplifies to $\Delta \Omega = \frac{\hbar}{4\pi R^2 m } \frac{1}{\sqrt{N}_t} \equiv \Delta \Omega_S$, where $N_t =\langle \hat{N}_+(0)\rangle$ is the total number of atoms. We take $\Delta \Omega_S$ as our benchmark sensitivity for the device. Importantly, the initial momentum distribution is irrelevant to the sensitivity, indicating that this sensitivity can be obtained regardless of the shape of the initial wave-packet. 

\begin{figure*}
\centering
\includegraphics[width=0.8\textwidth]{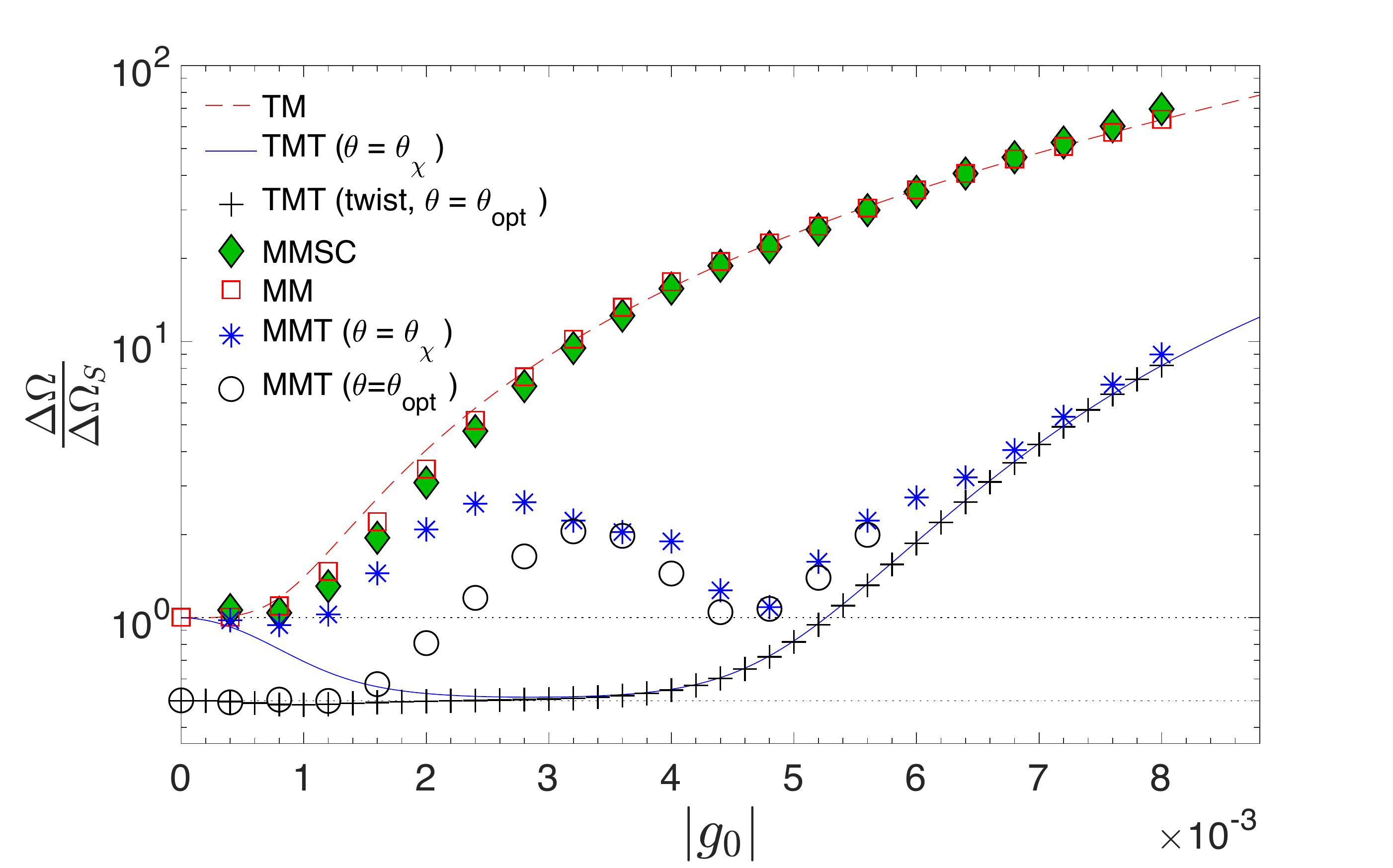}
\caption{(Color Online) Rotation sensitivity as a function of $g_0$ ($g_0$ is expressed in units of $\hbar^2/(mR)$). Red squares: Multi-mode TW (MMTW) model. Red dashed line: Analytic two-mode (TM) model. Green diamonds: multi-mode single-component (MMSC), Blue solid line: (Blue stars): Two-mode (multi-mode) TW model of pre-twisting scheme using $\theta =\theta_\chi$ (TMT, $\theta=\theta_\chi$ and MMT, $\theta=\theta_\chi$, respectively). Black circles (plus symbols): Multi-mode (two-mode) TW model of pre-twisting scheme with a numerically optimised $\theta = \theta_\mathrm{opt}$ for each point (TMT, $\theta=\theta_\mathrm{opt}$ and MMT, $\theta=\theta_\mathrm{opt}$, respectively). The upper black dotted line indicates $\Delta \Omega_S$, and the lower black dotted line indicates $\frac{1}{2}\Delta \Omega_S$, which is the standard sensitivity for matterwaves traversing two revolutions of the ring. Parameters: $N_t = 10^4$ and $k_0 = 80/R$ for all simulations, which corresponds to a maximum interaction parameter of $\chi T = 7.6\times 10^{-3}$. }
\label{fig1}
\end{figure*}

\subsection{Soliton Regime} To model the behaviour of our system in the soliton regime we choose the initial state $|\Psi(0)\rangle = \mathcal{D}(\ahat_+)\mathcal{D}(\ahat_-)|0\rangle$, where $\mathcal{D}(\ahat_\pm) = \exp (\alpha \ahat_\pm^\dag - \alpha^* \ahat_\pm)$ is the Glauber coherent displacement operator, $\alpha =\sqrt{N_s}$, where $N_s = N_t/2$ is the mean number of atoms in each mode, $\ahat_\pm = \int \Psi_\pm^*\psihat_\pm d \xi$, and $\Psi_\pm(\xi) = B \,\mathrm{sech} ( \sqrt{2 m |\mu | / \hbar^2} \xi) e^{\pm i k_0 \xi}$, where $k_0 = n/R$, and $B$ is a normalisation constant such that $\int_{-\pi R}^{\pi R} |\Psi_\pm|^2 d \xi = 1$.  The chemical potential $\mu$ is related to the number $N_s$ by $\mu = -N_s^2 g_0^2 m/8\hbar^2 $. We note that as we have started with our atoms already split between the two components, we forgo the first beamsplitter, allowing us to easily prepare the wave-packets with the correct shape for their occupation numbers. It was previously shown that the dynamics of such systems is reasonably insensitive to the \emph{total} population statistics, but \emph{is} sensitive to the statistics of the population \emph{difference} \cite{Haine:2009}. We chose a two-mode Glauber coherent state for our initial state as it reflects the number difference statistics that are obtained from coherent splitting of an ensembles of atoms.  Alternatively, we could have used a coherent spin state \cite{Radcliffe:1971}, which also has this property but for a well-defined total number of atoms. However a Glauber coherent state is much less computationally demanding for the numerical technique employed in this work. 

We simulate the dynamics of the system by using a stochastic phase space technique known as the Truncated Wigner (TW) method, which has previously been used to model the dynamics of quantum gasses \cite{Steel:1998, Sinatra:2002, Norrie:2006, Drummond:2017}, and unlike the GPE, can be used to model nonclassical particle correlations \cite{Ruostekoski:2013, Haine:2014, Szigeti:2017}. The derivation of the TW method has been described in detail elsewhere \cite{Drummond:1993, Steel:1998, Blakie:2008}. Briefly, the equation of motion for the Wigner function of the system can be found from the von-Neumann equation by using correspondences between differential operators on the Wigner function and the original quantum operators \cite{Gardiner:2004b}. By truncating third- and higher-order derivatives (the Truncated Wigner Approximation), a Fokker-Planck equation (FPE) is obtained. The FPE is then mapped to a set of stochastic partial differential equations for complex fields $\psi_j(\xi,t)$, which loosely correspond to the original field operators $\psihat_j(\xi,t)$, with initial conditions stochastically sampled from the appropriate Wigner distribution \cite{Blakie:2008, Olsen:2009}. The complex fields obey the partial differential equation
\begin{equation}
i\hbar\dot{\psi}_j = \left[\hat{H}_0 + g_0(|\psi_j|^2 -   \frac{1}{\Delta})\right]\psi_j \, ,\label{TWeq}
\end{equation}
where $\Delta$ is the element that characterises the discretisation of the spatial grid $\xi$. By averaging over many trajectories with stochastically sampled initial conditions, expectation values of quantities corresponding to symmetrically ordered operators in the full quantum theory can be obtained via the correspondence $\langle \{ f(\psihat^\dag_j, \psihat_j\}_\mathrm{sym}\rangle = \overline{f[\psi_j^*, \psi_j]}$, where `sym' denotes symmetric ordering and the overline denotes the mean over many stochastic trajectories. The initial conditions for the simulations are chosen as $\psi_\pm(\xi,0) = \sqrt{N_s}\Psi_\pm(\xi) + \eta_\pm(\xi)$, where $\eta_\pm(\xi)$ are complex Gaussian noises satisfying $\overline{\eta^*_m(\xi_i)\eta_n(\xi_j)} = \frac{1}{2}\delta_{m,n}\delta_{i,j}/\Delta$, for spatial grid points $\xi_i$ and $\xi_j$. Equations (\ref{TWeq}) was solved numerically on a spatial grid with $512$ points.

At $t=T$ the wave-packets have completed one circuit of the ring and a beam-splitter implemented via the transformation $\psi_{\pm} \rightarrow \frac{1}{\sqrt{2}}(\psi_{\pm} -i \psi_{\mp} e^{\pm 2i k_0\xi})$, before the expectation value and variance of the total number of particles in each component is calculated.  We calculate $\partial_\Omega \langle \hat{N}_d \rangle$ by using finite difference and simulating small variations of $\Omega$ around $\Omega = 0$. Fig.~(\ref{fig1}) (red squares) shows the rotation sensitivity as a function of the interaction strength $g_0$. We see that as $|g_0|$ increases, the sensitivity is rapidly degraded. We also analysed a single component system where the beam-splitting was performed by quantum reflection/transmission from a narrow barrier as in \cite{Helm:2015}. Fig.~(\ref{fig1}) (green diamonds) shows similar behaviour to the two-component system. For comparison, we have also modelled a noninteracting gas, for a variety of initial wave-packet with the same quantum statistics. For the two-component case, the sensitivity was equal to $\Delta \Omega_S$ in all cases. For the single component system, the sensitivity was also well approximated by $\Delta \Omega_S$ as long as the final state was still well approximated by two, well separated wave-packets. For gaussian wave-packets, this is achieved when $\sigma_\xi k_0 \gtrsim 1$ and $\sigma_\xi \lesssim \pi R$, where $\sigma_\xi$ is the initial width of the wave-packet. Outside this regime, the sensitivity decreased when the final width of the wave-packets was of the order of the circumference of the ring, and could no longer be distinguished from each other.  We note that making a measurement of the systems angular momentum, rather than position, may relax this constraint further.

\section{Two-mode model} The origin of this degradation is the quantum fluctuations in the population difference leading to uncertainty in the energy of each soliton, resulting in phase-fluctuations before the final beam-splitter. For small fluctuations in particle number $N$ around $N_s$, the energy of a single soliton is well approximated by 
\begin{equation}
E_N \approx E_{N_s} + \partial_{N_s} E_{N_s} (N-N_s) + \frac{1}{2}\partial^2_{N_s} E_{N_s} (N - N_s)^2,
\end{equation}
where $E_{N_s} \approx \left(\frac{\hbar^2 k_0^2 }{2m} \mp \Omega \hbar k_0R\right) N_s - \frac{g_0^2 m}{24\hbar^2}N_s^3$ is obtained by substituting $\psihat_{\pm} \rightarrow \sqrt{N}_s \Psi_{\pm}$ into \eq{ham1} and making the approximation that the limits of integration are $\pm \infty$. We can model the effect of the number fluctuations with an effective two-mode Hamiltonian \cite{Johnsson:2007a, Haine:2009, Riedel:2010, Haine:2011, Haine:2014, Kolar:2015} $\Ham = \Ham_{0} + \Ham_\mathrm{int}$, with
\begin{align}
\Ham_0 &= E_0 \sum_{j=+-} \ahatd_j\ahat_j- \hbar \Omega R k_0 \left(\ahatd_+\ahat_+ - \ahatd_-\ahat_-\right) \, , \\
\Ham_{\mathrm{int}} &= \frac{\hbar\chi}{2}\left(\ahatd_+\ahatd_+\ahat_+\ahat_+ + \ahatd_-\ahatd_-\ahat_-\ahat_- \right) \, ,
\end{align}
where $\hbar \chi = \partial^2_{N_s} E_{N_s} = - g_0^2 m N_s/4 \hbar^2$, and $E_0 =  \frac{\hbar^2k_0^2}{2m} + g_0^2 m N_s^2/8\hbar^2$. The form of $E_0$ is inconsequential as it results in a phase-shift that is common to both modes.  Moving to an interaction picture where the operators evolve under $\Ham_0$ and our state evolves under $\Ham_\mathrm{int}$, and expressing the state in the Fock basis gives
\begin{equation}
|\Psi(T)\rangle = e^{-|\alpha|^2}\sum_{n_1=0}^\infty\sum_{n_2=0}^\infty \frac{\alpha^{n_1}}{\sqrt{n_1!}}\frac{\alpha^{n_2}}{\sqrt{n_2!}}|n_1, n_2\rangle e^{-i\Phi_{n_1, n_2}} \, , \label{psi}
\end{equation}
where $\Phi_{n_1, n_2} = \frac{1}{2}\chi T \left[n_1(n_1-1) + n_2(n_2-1)\right]$. Introducing the psuedo-spin operators $\{\hat{J}_x = \frac{1}{2}(\ahat_+ \ahatd_- + \ahatd_+\ahat_-), \, \hat{J}_y = \frac{i}{2}(\ahat_+\ahatd_- - \ahatd_+\ahat_-), \, \hat{J}_z = \frac{1}{2}(\ahatd_+\ahat_+-\ahatd_-\ahat_-)\} = \frac{1}{2} \hat{N}_d$, at the final time $t=t_f$, evolution under $\Ham_0$ for a period $T$ followed by the final beamsplitter performs the transformation $\hat{J}_z(t_f) = -(\cos \phi_\Omega \hat{J}_y(0) +\sin \phi_\Omega \hat{J}_x(0))$.  At $\Omega =0$, \eq{delta_Omega} becomes $\Delta \Omega = \frac{\hbar}{4m\pi R^2}\sqrt{\mbox{Var}(\hat{J}_y)/\langle \hat{J}_x\rangle^2}$. Using \eq{psi}, we obtain
\begin{align}
\mbox{Var}(\hat{J}_y) &=\frac{N_t}{4} +  \frac{N_t^2}{8}\left(1 - \exp\left[-2N_t \sin^2(\chi T)\right] \right) \, , \\
\langle \hat{J}_x\rangle &= \frac{N_t}{2} \exp\left[N_t \left(-1 + \cos(\chi T)\right)\right] \, .
\end{align}
Fig.~(\ref{fig1}) (red dashed line) shows that our analytic model gives excellent agreement with both our single-component and two-component multi-mode numeric calculations. 

\begin{figure*}
\centering
\includegraphics[width=0.6\textwidth]{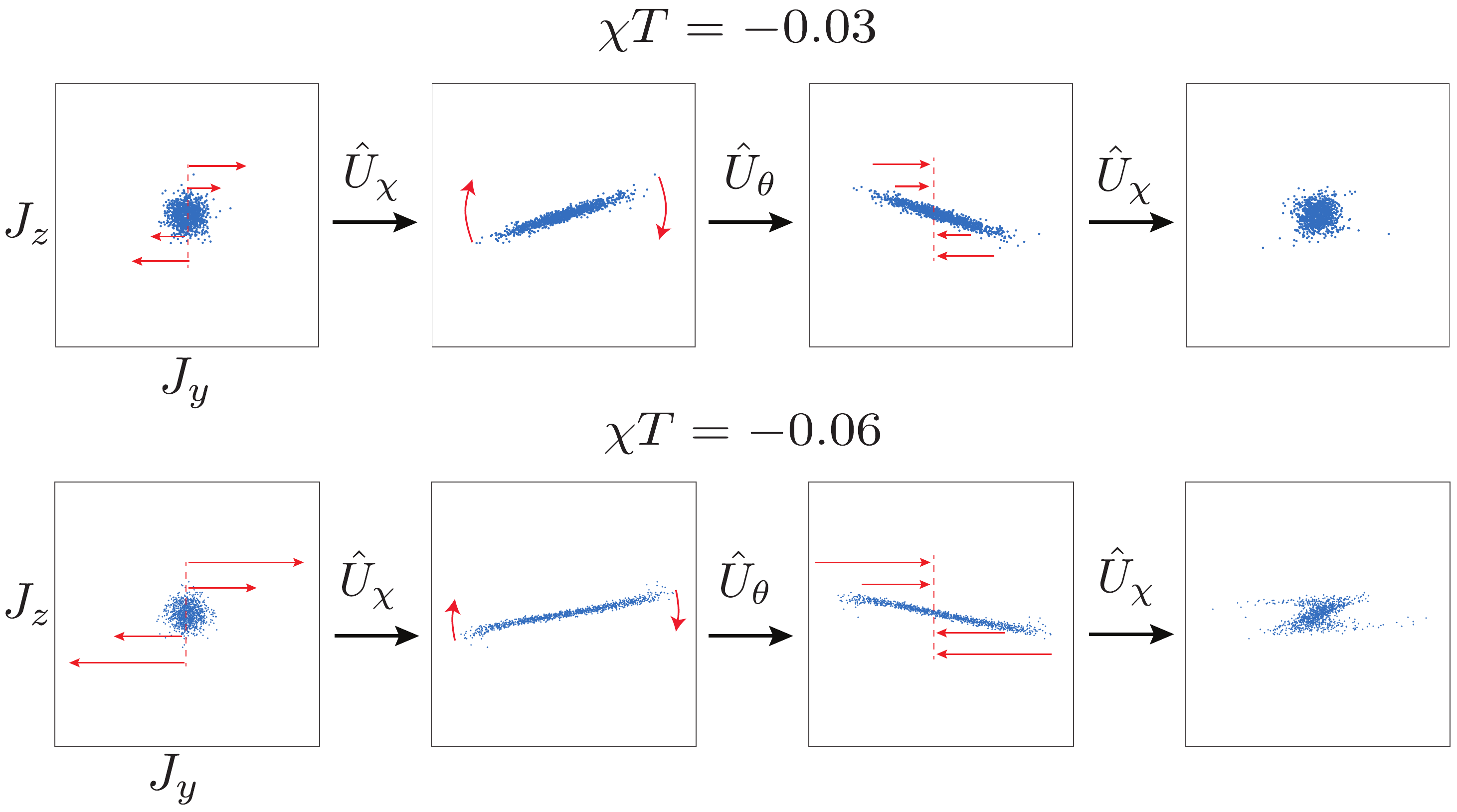}
\caption{(Color Online) Quasi-probability distribution for the pre-twisting sequence. Left to right: $|\Psi_0\rangle$, $\hat{U}_\chi|\Psi_0\rangle$, $\hat{U}_\theta\hat{U}_\chi|\Psi_0\rangle$, and $\hat{U}_\chi \hat{U}_\theta\hat{U}_\chi|\Psi_0\rangle$, for $\hat{U}_\chi = \exp(-i\Ham_{\mathrm{int}} T/\hbar)$. Top line: $\chi T = - 0.03$. Bottom line: $\chi T = - 0.06$. For visual clarity, a reduced number of atoms ($N_t = 100$) was used. }
\label{fig2}
\end{figure*}

\section{Pre-twisting to reduce the effects of phase diffusion}  
We can partially restore the sensitivity by implementing a pre-twisting scheme to reverse the effect of $\Ham_\mathrm{int}$. Fig.~(\ref{fig2}) shows a quasi-probability distribution formed from individual trajectories from a 2-mode TW simulation evolving under $\Ham_\mathrm{int}$. Initially, the individual trajectories are spread out in both $\hat{J}_z$ and $\hat{J}_y$. However, after a period of evolution, the spread in $\hat{J}_z$ is converted into a much larger spread in $\hat{J}_y$, which is the origin of the degradation. By applying a rotation $\hat{U}_\mathrm{\theta} = e^{i \theta \hat{J}_x}$, the state is twisted such that a second period of evolution under $\Ham_\mathrm{int}$ approximately revives the initial state. However, this process breaks down for larger values of $\chi T$, as can be seen in the lower panels of fig.~(\ref{fig2}). This is because for small values of $\chi T$, the trajectories roughly form an ellipse, which when rotated, is similar in shape to its reflection about the $\hat{J}_z$ axis. However, for larger values of $\chi T$, the trajectories form a \emph{bent} ellipse, which when rotated about the $\hat{J}_x$ axis, deviates significantly from its reflection about the $\hat{J}_z$ axis, and thus the second period of nonlinear evolution does not revive the initial state \cite{Nolan:2017b, Mirkhalaf:2018}. We note that this process could also have been achieved by simply reversing the sign of $\chi$ for the second period of evolution. However, this is incompatible with the use of bright solitons as the require a negative interaction constant.  The rotation angle that performs the rotation illustrated in fig.~(\ref{fig2}) is 
\begin{equation}
\theta_\chi = -\cos^{-1} (-\gamma), \label{eq_theta_chi}
\end{equation}
with 
\begin{equation}
\gamma = \frac{\left(\exp(2 s N_t) -1\right)}{\sqrt{ \left(\exp(2sN_t) -1\right)^2 + 16 s \exp\left(2N_t(\cos \chi T - \cos 2\chi T)\right)}},
\end{equation} 
and $s = \sin^2 \chi T$, and is derived in appendix (\ref{sec_appendix}).

The sensitivity that this scheme provides is shown in fig.~(\ref{fig1}) (blue solid line). There are two factors that influence the sensitivity. The first is the reduction in quantum noise ($\Var(\hat{J}_y)$) due to this pre-twisting scheme. The second is that the $\theta$ rotation has a non-trivial effect on $d \langle \hat{J}_z\rangle /d \Omega$ due to the interplay between the phase shift accumulated before and after the twisting, with $\theta =0 (\pi)$ leading to perfect addition(cancellation) of this phase.  As such, for small values of $\chi T$, $\theta_\chi$ is not the optimum angle, as the reduction in variance is offset by the partial cancellation of phase accumulation. To obtain higher sensitivities, we optimise $\theta$ numerically. The optimum sensitivity is shown in fig.~(\ref{fig1}) (black crosses). The optimum actually dips slightly below the standard quantum limit (SQL) because the final state in this case has reduced fluctuations in $\Var(\hat{J}_y)$. 

We implement the pre-twisiting scheme in our multi-mode model by replacing the final 50/50 beamsplitter of the single loop scheme with a variable angle beam splitter performing the transformation $\psi_\pm \rightarrow  \psi_\pm \cos \theta - i  \psi_\mp\sin\theta e^{\pm 2i k_0 \xi}$, and then allowing the solitons to perform a second circuit of the ring before the final 50/50 beamsplitter is implemented. Again, such a transformation is easily implemented via a coherent two-photon Raman transition. However, when assessing the performance of this scheme (fig.~(\ref{fig1}) blue stars), we see that while there is generally some improvement in sensitivity when compared to the original scheme, there is a significant discrepancy between the 2-mode model and the multi-mode model. In particular, the multi-mode model predicts significantly worse sensitivity than compared to the two-mode model for intermediate values of $|g_0|$. For larger values of $|g_0|$, the multi-mode model still gives about an order of magnitude improvement compared to the original scheme, but this is still worse than what would be obtained by using a non-interacting gas. In an attempt to further improve the sensitivity, we numerically optimised $\theta$ (fig.~(\ref{fig1}) black circles). This results in significant improvement for small values of $|g_0|$. However, the `bump' for intermediate values of $|g_0|$ is still present. We speculate that the origin of this behaviour is different regions of the wave-packet experiencing different degrees of phase shearing. This is noticible when the pre-twisting is attempted, as the multiple regions would require slightly different rotations angles to perfectly revive the state - which is a requirement our pre-twisting scheme is not capable of. We also attempted to implement the pre-twisting scheme in the single-component system by varying the height of the barrier to implement $\hat{U}_\theta$. However, we found very little improvement compared to the single-loop scheme. We suspect that this is partly due to the difficulty in controlling $\theta$ and the phase of the outgoing matterwaves after interaction with the barrier. 

\section{Discussion} Our results generally indicate that for the case of rotation sensing with a two-component system, it is better to work in a regime with minimal interactions rather then pursuing the use of bright solitons. If working in a regime where interactions are unavoidable, then one should consider using the pre-twisting scheme presented in this letter. In a single component system, minimising interactions and ensuring that the wave-packet satisfies the conditions for distinguishable wave-packets, is favourable to the use of bright solitons. In situations when these conditions cannot be met, it may be the case that bright solitons provide superior performance. As the sensitivity scales with the enclosed area of the device, it is beneficial to increase the circumference of the ring. However, when working in the soliton regime, assuming the magnitude of the momentum kick is held fixed, the time taken for the solitons to complete a circuit, and therefore the amount of phase diffusion, increases with the size of the ring. This will ultimately limit the obtainable sensitivity. In the linear regime however, the expansion of the wave-packets scales linearly with time, such that the conditions for wave-packet distinguishability is approximately independent of the ring circumference (the fraction of the circumference covered by each wave-packet at the final time is independent of the circumference), so no such limitations exist. 

As the phase-diffusion mechanism investigated in this manuscript will also be present in any sensing schemes involving bright-solitons, the results of this paper suggest that one should always use models that include quantum noise rather than relying exclusively on mean-field models to assess the metrological sensitivity. 

However, we do not claim that the use of bright solitons is entirely without benefit. Wave-packet spreading may prove problematic if beamsplitters that transfer linear momentum (rather than angular momentum, as considered in this paper) are used, as a spatially non-localised source will experience a radial component to the momentum transfer, causing mode-matching issues. Additionally, it may be possible that some detection systems are less susceptible to imperfections if the matterwaves remain spatially localised. It was observed in the experiment of McDonald \etal \cite{McDonald:2014} that the maximum sensitivity was achieved when the scattering length was tuned to create a soliton. The reason for this was likely that the reduction in dispersion reduced various sources of technical noise such as imperfections in the trapping potential. Furthermore, for the interrogation times used, the two soliton wave-packets remained spatially overlapping for the duration of the experiment, so the system would not be subjected to the relative phase shearing noise reported in this manuscript. Additionally, the experiment was not operating at the SQL so it is unlikely that this noise source would be observed. 

Finally, we note that it has been shown that soliton dynamics can create non-classical states \cite{Weiss:2009, Streltsov:2009, Lewenstein:2009, Martin:2012, Gertjerenken:2013}. However, it has yet to be shown that these states can be used for enhanced matterwave interferometry, as they will be subject to the same phase diffusion which is the subject of this manuscript, and further modelling of these systems should be pursued. 

\section{Acknowledgements} The author would like to acknowledge useful discussions with Samuel Nolan, Matthew Davis, Joel Corney, Murray Olsen, Stuart Szigeti, Michael Bromley, John Helm, Simon Gardiner, and Nick Robins. The numerical simulations were performed with XMDS2 \cite{Dennis:2013} on the University of Queensland School of Mathematics and Physics computing cluster ``Dogmatix'', with thanks to Ian Mortimer for computing support. This work was supported by the European Union's Horizon 2020 research and innovation programme under the Marie Sklodowska-Curie grant agreement No.~704672.

\bibliography{../../simon_bib}

\begin{thebibliography}{62}
\bibitem [{not()}]{note1}%
  \BibitemOpen
  \href@noop {} {}\bibinfo {howpublished} {The only effect of $g_{+-}$ is a
  slight modification of the velocity of the wavepackets, slightly altering the
  collision time $T$.}\BibitemShut {Stop}%
\end{thebibliography}

\newpage

\begin{widetext}
\section*{Supplemental Material: Quantum Noise in Soliton Matter-Wave Interferometry}
In this supplemental material we provide further details on the calculations in the main text. Specifically, we derive the rotation angle required for the pre-twisting scheme. 

\section{Derivation of $\theta_\chi$ (Eq (\ref{eq_theta_chi}))} \label{sec_appendix}
In this appendix, we provide further details on the calculations in the main text. Specifically, we derive the rotation angle required for the pre-twisting scheme. The angle required for our pre-twisting scheme, $\theta_\chi$, is the angle such that rotation about the $J_x$ axis returns the variance of $J_z$ to its original value, as illustrated in Fig.~\ref{suppfig1}. 

\begin{figure}
\includegraphics[width=0.4\columnwidth]{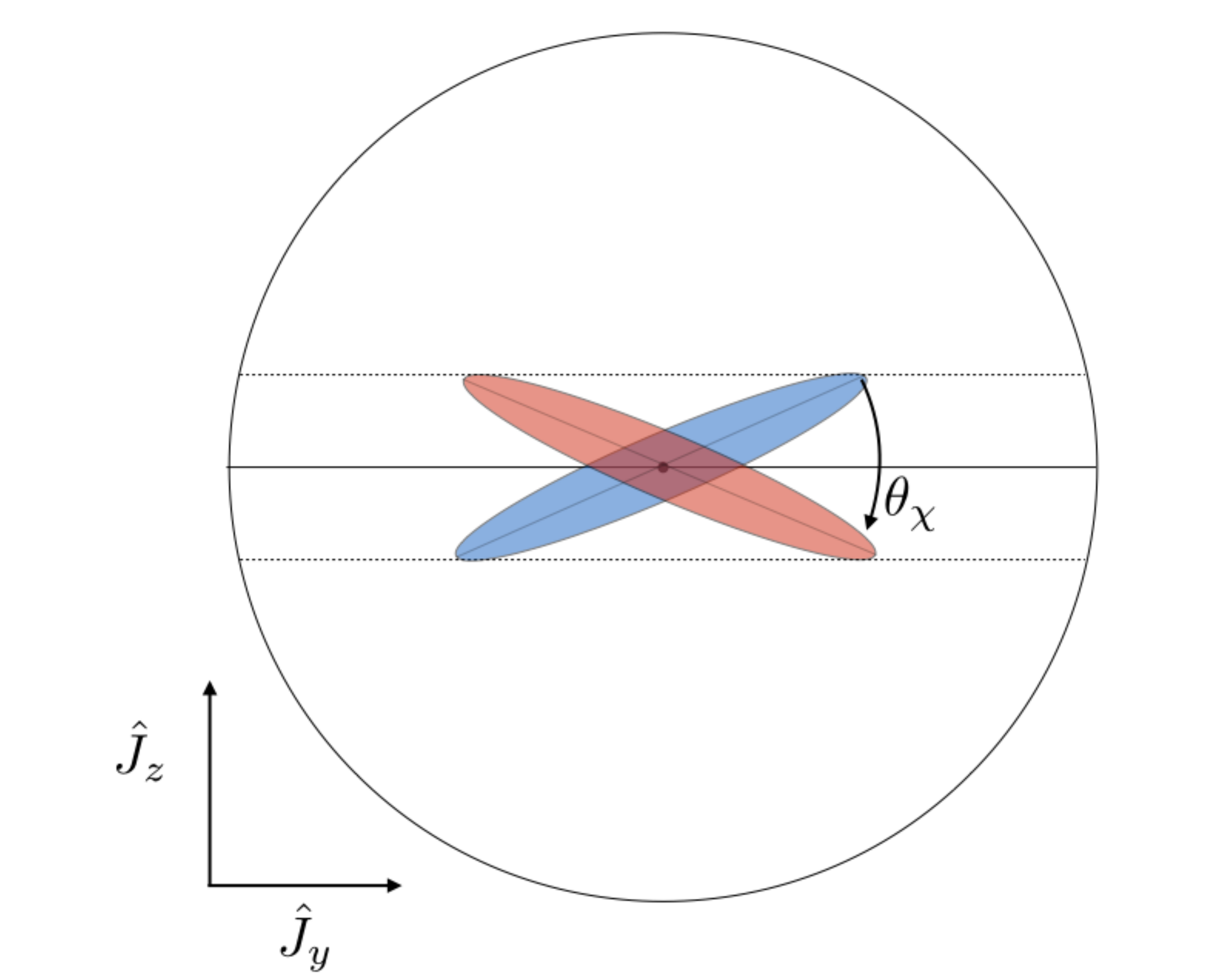}
\caption{(Color Online) The rotation angle $\theta_\chi$ about the $J_x$ axis required for re-phasing after a second application of $\hat{U}_\chi$ is the angle that has the same variance in $\hat{J}_z$ before and after rotation.   }
\label{suppfig1}
\end{figure}

The action of the variable angle beamsplitter $\hat{U}_\theta = \exp ( i \theta_\chi \hat{J}_x)$ on the psuedo-spin operators before ($t=T$) and after ($t=t_1$) the rotation is
\begin{align}
\hat{J}_z(t_1) &= \hat{U}_\theta^\dag \hat{J}_z(T) \hat{U}_\theta =  \hat{J}_z(T) \cos \theta_\chi + \hat{J}_y(T) \sin \theta_\chi \\
\hat{J}_y(t_1) &= \hat{U}_\theta^\dag \hat{J}_y(T) \hat{U}_\theta =\hat{J}_y(T) \cos \theta_\chi - \hat{J}_z(T) \sin \theta_\chi \, .
\end{align}
Therefore, after the rotation, the variance in $\hat{J}_z$ is
\begin{align}
V(\hat{J}_z(t_1)) &= \langle \hat{J}^2_z(t_1)\rangle \notag \\
&= \langle \hat{J}^2_z(T)\rangle \cos^2\theta_\chi + \langle \hat{J}^2_y(T)\rangle \sin^2\theta_\chi  \notag \\ 
&+ \cos\theta_\chi\sin\theta_\chi \left(\langle \hat{J}_z(T)\hat{J}_y(T)\rangle + \langle \hat{J}_y(T)\hat{J}_z(T)\rangle\right) \, ,
\end{align}
since the state is chosen such that $\langle \hat{J}_z(t_1)\rangle = \langle \hat{J}_z(T)\rangle = 0$. The evolution under $\hat{U}_\chi = \exp (-i \Ham_\mathrm{int}T)$ commutes with $\hat{J}_z$, so $\langle \hat{J}^2_z(T)\rangle = \langle \hat{J}^2_z(0)\rangle = N_t/4$. The angle $\theta_\chi$ is defined as the angle such that $\langle \hat{J}_z^2(t_1)\rangle = \langle \hat{J}_z^2(T)\rangle = N_t/4$. Expressing the pseudo-spin operators in terms of bosonic creation and annihilation operators, and making the substitution $\ahat_+ \rightarrow \ahat$ and $\ahat_- \rightarrow \bhat$ for ease of notation gives
\begin{align}
\hat{J}_z^2 &= \frac{1}{4}\left( \ahatd\ahatd\ahat\ahat + \bhatd\bhatd\bhat\bhat + \ahatd\ahat + \bhatd\bhat - 2\ahatd\ahat\bhatd\bhat\right) \\
\hat{J}_y^2 &= \frac{1}{4}\left( 2\ahatd\ahat\bhatd\bhat + \ahatd\ahat + \bhatd\bhat
- \ahatd\ahatd\bhat\bhat - \bhatd\bhatd\ahat\ahat \right) \\
\hat{J}_z\hat{J_y} + \hat{J}_y\hat{J}_z &= \frac{i}{2}\left(\ahatd\ahat\ahat\bhatd + \ahatd\bhatd\bhat\bhat  - \ahat\bhatd\bhatd\bhat -\ahatd\ahatd\ahat\bhat \right)
\end{align}
In order to evaluate these expressions,  we need to calculate terms such as $\langle \ahatd\ahatd\ahat\bhat\rangle$ with respect to the state
\begin{equation}
|\Psi(T)\rangle = e^{-|\alpha|^2}\sum_{n_1=0}^\infty\sum_{n_2=0}^\infty \frac{\alpha^{n_1}}{\sqrt{n_1!}}\frac{\alpha^{n_2}}{\sqrt{n_2!}}|n_1, n_2\rangle e^{-i\Phi_{n_1, n_2}} \, , \label{psi_supp}
\end{equation}
where
\begin{equation}
\Phi_{n_1, n_2} = \frac{1}{2}\chi T \left[n_1(n_1-1) + n_2(n_2-1)\right] \, ,
\end{equation}
and $\alpha = \sqrt{N_t/2}$. We will explicitly compute one example, and provide the rest of these operator moments in a table. 
\begin{align}
\langle \ahatd\ahatd\ahat\bhat\rangle &= e^{-2|\alpha|^2}\sum_{m_1=0}^\infty\sum_{m_2=0}^\infty\sum_{n_1=0}^\infty\sum_{n_2=0}^\infty\frac{(\alpha^*)^{m_1}(\alpha^*)^{m_2}\alpha^{n_1}\alpha^{n_2}}{\sqrt{m_1!m_2!n_1!n_2!}}\langle m_1, m_2|  \ahatd\ahatd\ahat\bhat|n_1, n_2\rangle e^{i(\Phi_{m_1, m_2}-\Phi_{n_1, n_2})} \notag \\
&= e^{-2|\alpha|^2}\sum_{m_1=0}^\infty\sum_{m_2=0}^\infty\sum_{n_1=1}^\infty\sum_{n_2=1}^\infty\frac{(\alpha^*)^{m_1}(\alpha^*)^{m_2}\alpha^{n_1}\alpha^{n_2}}{\sqrt{m_1!m_2!n_1!n_2!}} n_1\sqrt{n_1+1} \sqrt{n_2}  \langle m_1, m_2|n_1+1, n_2-1\rangle e^{i(\Phi_{m_1, m_2}-\Phi_{n_1, n_2})} \notag \\
&= e^{-2|\alpha|^2}\sum_{m_1=0}^\infty\sum_{m_2=0}^\infty\sum_{n_1=1}^\infty\sum_{n_2=1}^\infty\frac{(\alpha^*)^{m_1}(\alpha^*)^{m_2}\alpha^{n_1}\alpha^{n_2}}{\sqrt{m_1!m_2!n_1!n_2!}} n_1\sqrt{n_1+1} \sqrt{n_2} e^{i(\Phi_{m_1, m_2}-\Phi_{n_1, n_2})}\delta_{m_1, n_1+1}\delta_{m_2, n_2-1} \notag \\
&= e^{-2|\alpha|^2}\sum_{m_2=0}^\infty\sum_{n_1=1}^\infty\frac{(\alpha^*)^{n_1+1}(\alpha^*)^{m_2}\alpha^{n_1}\alpha^{m_2+1}}{\sqrt{((n_1+1)!m_2!n_1!(m_2+1)!}} n_1\sqrt{n_1+1} \sqrt{m_2+1} e^{i\chi T(n_1-m_2)} \notag \\
&= |\alpha|^4 e^{i\chi T} e^{-2|\alpha|^2}  \sum_{m_2=0}^\infty\sum_{n_1=1}^\infty \frac{(|\alpha|^2 e^{i\chi T})^{n_1-1}}{(n_1-1)!}\frac{(|\alpha|^2 e^{-i\chi T})^{m_2}}{m_2!} \notag \\
&= |\alpha|^4 e^{i\chi T} e^{-2|\alpha|^2} e^{|\alpha|^2 e^{i\chi T}}e^{|\alpha|^2 e^{-i\chi T}} \notag \\
&= \frac{N_t^2}{4} e^{i\chi T} \exp \left( N_t(\cos \chi T - 1)\right) \, .
\end{align}
The complete set of moments required to calculate $\langle \hat{J}^2_z(t_1)\rangle$ is
\begin{subequations}
\begin{align}
\langle \ahatd \ahat\rangle & = \langle \bhatd\bhat \rangle = \frac{N_t}{2} \\
\langle \ahatd\ahatd \ahat\ahat\rangle & = \langle \bhatd\bhatd \bhat\bhat \rangle = \frac{N_t^2}{4} \\
\langle \ahatd\ahatd \bhat\bhat\rangle &= \langle \ahat\ahat \bhatd\bhatd\rangle =\frac{N_t^2}{4}\exp \left[N_t (\cos 2\chi T-1)\right] \\
\langle \ahatd\ahatd\ahat\bhat\rangle &= \langle \ahat \bhatd\bhatd\bhat\rangle = \frac{N^2_t}{4}e^{i\chi T} \exp\left[N_t(\cos \chi T-1)\right] \\
\langle \ahatd\ahat\ahat\bhatd\rangle &= \langle \ahatd \bhatd\bhat\bhat\rangle = \frac{N^2_t}{4}e^{-i\chi T} \exp\left[N_t(\cos \chi T-1)\right] \, .
\end{align}
\end{subequations}
The solution to $\langle \hat{J}_z^2(t_1)\rangle = \langle \hat{J}_z^2(T)\rangle$ for $\theta_\chi$ gives four non-trivial solutions:
\begin{subequations}
\begin{align}
\theta_\chi &= \cos^{-1} (\gamma) \\
\theta_\chi &= \cos^{-1} (-\gamma) \\
\theta_\chi &= -\cos^{-1} (\gamma) \\
\theta_\chi &= -\cos^{-1} (-\gamma) \, ,\label{theta_sol}
\end{align}
\end{subequations}
where 
\begin{align}
\gamma &= \frac{\langle \hat{J}^2_z(T)\rangle - \langle \hat{J}^2_y(T)\rangle}{\sqrt{\langle \hat{J}^2_z(T)\rangle^2 + \langle \hat{J}^2_y(T)\rangle^2 + \langle \hat{J}_z(T)\hat{J}_y(T) + \hat{J}_y(T)\hat{J}_z(T)\rangle^2 - 2\langle \hat{J}_z^2(T)\rangle\langle \hat{J}^2_y(T)\rangle }} \, \notag \\
&= \frac{\left(\exp(2N_t\sin^2\chi T) -1\right)}{\sqrt{ \left(\exp(2N_t\sin^2\chi T) -1\right)^2 + 16\sin^2\chi T \exp\left(2N_t(\cos \chi T - \cos 2\chi T)\right)}} \, .
\end{align}
Of those solutions, the only one that gives better performance than the single loop scheme is \eq{theta_sol}, which is what was used for both the two-mode and multi-mode pre-twisting calculations. 
\end{widetext}

\end{document}